# Polytypism of Incommensurately Modulated Structures of Crystalline Bromine upon Molecular Dissociation under High Pressure

Yuqing Yin[1,2*], Andrey Aslandukov[1,3], Maxim Bykov[4], Dominique Laniel[5], Alena Aslandukova[3], Anna Pakhomova[6], Timofey Fedotenko[7], Wenju Zhou[1], Fariia Iasmin Akbar[1,3], Michael Hanfland[6], Konstantin Glazyrin[7], Carlotta Giacobbe[6], Eleanor Lawrence Bright[6], Gaston Garbarino[6], Zhitai Jia[2], Natalia Dubrovinskaia[1,8], Leonid Dubrovinsky[3*]

**Affiliations:**
[1]Material Physics and Technology at Extreme Conditions, Laboratory of Crystallography, University of Bayreuth, 95440 Bayreuth, Germany
[2]State Key Laboratory of Crystal Materials, Shandong University, Jinan 250100, China
[3]Bayerisches Geoinstitut, University of Bayreuth, 95440 Bayreuth, Germany
[4]Institute of Inorganic Chemistry, University of Cologne, Greinstrasse 6, 50939 Cologne, Germany
[5]Centre for Science at Extreme Conditions and School of Physics and Astronomy, University of Edinburgh, EH9 3FD Edinburgh, United Kingdom
[6]European Synchrotron Radiation Facility, B.P.220, F-38043 Grenoble Cedex, France
[7]Photon Science, Deutsches Elektronen-Synchrotron, Notkestrasse 85, 22607 Hamburg, Germany
[8]Department of Physics, Chemistry and Biology (IFM), Linköping University, SE-581 83, Linköping, Sweden
*Corresponding authors. Email: Yuqing.Yin@uni-bayreuth.de (Y.Y.); Leonid.Dubrovinsky@uni-bayreuth.de (L.D.)

**Abstract:**

Polytypism of incommensurately modulated structures was hitherto unobserved. Here, we found the phenomenon in simple halogen systems of bromine and iodine upon molecular dissociation in the solids under pressure. Single-crystal synchrotron X-ray diffraction in laser heated diamond anvil cells pressurised up to 112 GPa revealed a number of allotropes of bromine and iodine including polytypes of Br-III$\gamma$ (*Fmmm*(00$\gamma$)*s*00) with $\gamma$ varying within 0.18 to 0.3.

Studies of the behavior of halogens under pressure provide an insight into how simple molecular systems respond to extreme conditions. Fundamental problems to be addressed are the structure evolution [1], the band gap closure and metallization [2-4], phase transformations during molecular dissociation [4,5], and others. Extensive theoretical studies based on density functional theory (DFT) calculations [6-9] and experimental investigations using powder X-ray diffraction (XRD) [4,5], X-ray absorption spectroscopy (XAS) [1,10], and Raman spectroscopy measurements [4,11,12] have been carried out to determine the behavior of halogens under pressure. In fact, experimental studies have become especially important, as DFT calculations face challenges in application to halogen systems [13,14].

There are experimental observations suggesting similar phase transitions in chlorine, bromine, and iodine, but at different pressures (Fig. 1). This is consistent with the empirical rule of high-pressure crystal chemistry [15]: elements behave at high pressures like the elements below them in the periodic table at lower pressures. Iodine at ambient conditions is solid and has the $oC8$ molecular structure (called also I$_2$-I; space group *Cmce*), which is preserved up to 16 GPa [16]. (Fig. 1, Fig. S1a). Br$_2$ and Cl$_2$ solidify at ~0.5 GPa [17] and 1.15 GPa [4], respectively. These solid phases, *Cmce* Br$_2$-I [17] and *Cmce* Cl$_2$-I [4], possess the $oC8$ crystal structures similar to that of I$_2$-I [16]. These structures are preserved up to 82 GPa [18] and 223 GPa [4], respectively. Here and further the numbering of halogens' allotropes is adopted from the original papers.

Further compression modifies the intra- or intermolecular distances that results in a series of phase transformations, as presented in Fig. 1. At first, the molecular character of the structures of allotropes of chlorine (Cl$_2$-I´) and iodine (I$_2$-I´) is preserved. The structure of $C2/m$ Cl$_2$-I´ ($mC8$) [4] (Fig. 1, Fig. S1b) was suggested based on powder XRD, as well as that of $C2/m$ I$_2$-I´ ($mC8$) [19]. However, the latter was not confirmed by the recent single-crystal X-ray diffraction (SCXRD) studies of iodine [16], which showed that at ~16 GPa I$_2$-I transforms into the orthorhombic phase $Cmc2_1$ I$_2$-VI ($oC8$) [16] (Fig. 1, Fig. S1c), so that I$_2$-I´ [19] is not shown in Fig. 1. Hitherto, there has not been XRD data available for high-pressure bromine above 88 GPa [18], but the molecular phase, next after Br$_2$-I, was suggested to form above ~25 GPa based on XAS [1] measurements and DFT calculations [6]. Fig. 1 shows exclusively the phases for which structural information is available from experimental XRD data.

The pathways of crystalline chlorine and iodine from the molecular phases to the non-molecular (atomic) ones, *Immm* Cl-II ($oI2$) and *Immm* I-II ($oI2$) (Fig. S1f) go through phases with incommensurately modulated structures (further called incommensurate and designated by the letter *i* in front of their Pearson symbols) [4,5,16,20,21]. First structural information for an incommensurate phase of iodine was reported in 2003 by Takemura *et al.* [5] based on powder XRD data. Later, SCXRD (see Bykova *et al.* [16]) showed that within the same pressure interval there are two iodine phases with incommensurate orthorhombic $i$-$oF4$ structures: $Fmmm(00\gamma)s00$ I$_2$-VII ($\gamma = 0.4837$ at 20.8 GPa) at 20.8-24 GPa (Fig. S1d) and $Fmmm(00\gamma)s00$ I-V ($\gamma = 0.274$ at 24 GPa) at 24-32 GPa (Fig. S1e) (I$_2$-VII turned to be the same as the phase called I-VI in Ref. [20]; see Ref. [16] for details). The incommensurate phase $Fmm2(\alpha00)0s0$ Cl-V (*i-oF*4) was reported for chlorine in the pressure range of 223-258 GPa from powder XRD [4]. For bromine, a modulated phase similar to I-V was suggested to form above 80 GPa [12], but the experimental structural data are absent. Theoretical calculations predicted several structures for bromine (*oF*20, *oC*24, *oF*28, and *oC*12) through the commensurate approximation of the *i-oF*4 phase [7,13].

For bromine, accurate structural information at high pressures is still missing, and SCXRD data are needed to evaluate the path of its solid-state molecular dissociation. In this work, we present the results of synchrotron SCXRD experiments on solid bromine synthesised in diamond anvil cells (DACs) in the pressure range of 45(3) to 112(3) GPa after the laser heating of a solid precursor to ~2600 K.

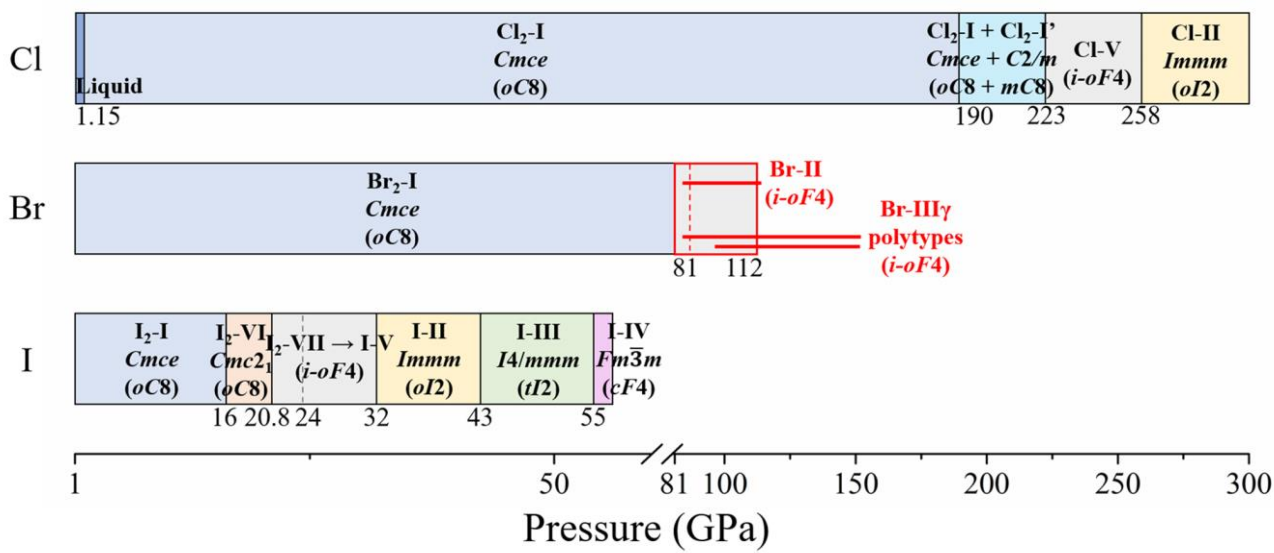

**Fig. 1. Schematic diagram presenting experimentally observed high-pressure allotropes of halogens, Cl [4], Br ([17,18] and this work - highlighted in red), and iodine [16,21-23], for which structural information was obtained from XRD data.** The allotropes are designated by Roman numbers (as adopted in the literature), and space groups and Pearson symbols (in brackets) are provided. The letter *i* added to the Pearson symbol (*i-oF*4) refers to incommensurately modulated structures with the space group $Fmmm(00\gamma)s00$. Phases indicated with the same color possess the same structure type. Crystal structures of all phases shown in the diagram are provided in Fig. S1.

Experiments were carried out in the pressure range of 45(3) to 112(3) GPa using the BX90-type screw-driven DACs [24] (see Supplementary Methods and Table S1 for details). CBr$_4$ (Sigma-Aldrich, >99% purity) was used as a source of bromine because carbon tetrabromide decomposes at pressures over 40 GPa under heating and releases bromine [25-27]. As far as in all experiments in DACs carbon from diamond anvils is a part of the system, for it can diffuse during laser heating [28], the use of CBr$_4$ does not introduce any additional elements into the pressure chamber. At the same time, this precursor is much easier to load into a DAC than pure bromine.

The samples were first compressed to the desired pressures. SCXRD measurements were made on temperature-quenched samples after their laser-heating up to ~2600 K (see Table S1 for experimental details). At each pressure point high-quality diffraction spots (peaks are intense and relatively sharp) were detected. The obtained data were analyzed using our methodology for processing SCXRD from microcrystalline multiphase samples [29-31] with the implementation of the DAFi program [32].

In the main experiment with CBr$_4$ as a single precursor, SCXRD data were collected at 72(3), 81(3), 96(3), 105(3), and 112(3) GPa. Solid bromine *Cmce* Br$_2$-I was found to preserve its *oC*8 molecular structure at 72(3) GPa and room temperature (see Table S2 for crystallographic data). The analysis of SCXRD did not show any sign of co-existence of other molecular phases with *Cmce* Br$_2$-I (Fig. S2), unlike reports for Cl$_2$-I' [4] and I$_2$-I' [19].

Laser heating of the sample at 81(3) GPa led to the formation of two different kinds of crystalline domains: some with lattice parameters similar to those of *Cmce* Br$_2$-I (*oC*8), and others with different lattice parameters (slightly smaller $a$ (~4.8 vs ~4.6 Å), slightly larger $b$ (~3.3 vs ~3.5 Å), and essentially different $c$ (~7.5 vs ~14.3, 24.8, or 53.1 Å). The analysis of the diffraction data from the first kind of domains revealed clear splitting of some diffraction spots detected on the reconstructed reciprocal space images (Fig. S3). The diffraction pattern could be described by an orthorhombic *F*-centered unit cell and a modulation wavevector $\mathbf{q} = \gamma \mathbf{c}^*$, where $\gamma$ slightly deviates from 0.5 (Fig. S3). While a few of the 2$^{nd}$ order satellites could be observed, the average $F^2/\sigma(F^2)$ values for these reflections were low (~81, ~58, and ~13 for the main reflections, the 1$^{st}$ order, and the 2$^{nd}$ order satellites, respectively), and only the 1$^{st}$ order satellites were used for final refinements (a detailed comparison of the two refinements can be found in Table S3). Therefore, we accept here the structural model as the bromine phase II (Br-II) at 81(3) GPa, described by the (3+1)-dimensional superspace group $Fmmm(00\gamma)s00$ with structural parameters $a$ = 3.2767(14) Å, $b$ = 4.7924(19) Å, $c$ = 3.7618(16) at Å, $\gamma$ = 0.493(3), and $A_1(x)$ = 0.0742(11) (Table S3, CIF deposited at CSD 2289479). Displacive modulations of Br atom ($u_i(\overline{x_4})$ for $i = x, y, z$) are described by a truncated Fourier series, which take into account only the 1$^{st}$ order harmonics and due to the symmetry restrictions get the following form[33]: $u_x(\overline{x_4}) = A_1(x)\sin(2\pi\overline{x_4})$, $u_y = u_z = 0$ where $\overline{x_4} = t + \mathbf{q}\overline{x}$ ($t$ is the phase of the modulation).

The second type of domains at 81(3) GPa has the same structural motif as Br-II (*Fmmm*(00γ)*s*00), but with different γ values. Processing SCXRD data reveals the coexistence of bromine single-crystal domains with similar but distinct values of γ (between ~0.25 and ~0.30 at 81 GPa, for example), which we call Br-IIIγ and consider as polytypes (see below the discussion on the use of the term "polytypes"). Polytypism of the incommensurate structures is present across a wide pressure range (from 81(3) to 112(3) GPa) and reflects the way how the bromine structure transforms during molecular dissociation. Some polytypes of Br-IIIγ were selected as examples and deposited in the Cambridge Crystallographic Data Center (CCDC) under the identifiers CSD 2289480 – CSD 2289493. Their crystallographic data are provided in Tables S4-S8 (grouped by pressure), while Table S9 shows all bromine allotropes synthesized in the pressure range of 72(3) to 112(3) GPa in this work. Note that, like in the SCXRD studies of iodine [16], all satellite reflections of Br-IIIγ up to $2^{nd}$ order can be well described by an *oF*4 basic cell with only one modulation vector (Fig. S4, Tables S4-S8).

Crystal chemistry and chemical bonding characteristics of polytypes Br-II and Br-IIIγ are different. In incommensurate Br-II, the modulation of the Br-Br distances results in the structure that can be presented as a mixture of $Br_2$ molecules (at 81 GPa, taking $Br_2$-I as a reference (Fig. 2a), intramolecular distances are shorter than ~2.35 Å, and interatomic – longer than ~2.45 Å) and $Br_3$ zigzag chains (all distances between atoms in the chain are shorter than ~2.4 Å) (Fig. 2b). The Br-IIIγ polytypes do not contain $Br_2$ molecules (shortest Br-Br distances are of ~2.38 Å), but consist of nearly linear $Br_3$, $Br_4$, and $Br_5$ atomic chains (angles between atoms in the groups vary from ~177.5 to 180 degrees), and V-shape ($Br_7$) atomic chains (Fig. 2c). The structure of Br-II can be considered as intermediate between the structures of $Br_2$-I and Br-IIIγ. Unlike for iodine [16], we do not see any evidence of additional molecular phases (like the intermediate $I_2$-VI, space group *Cmc*$2_1$).

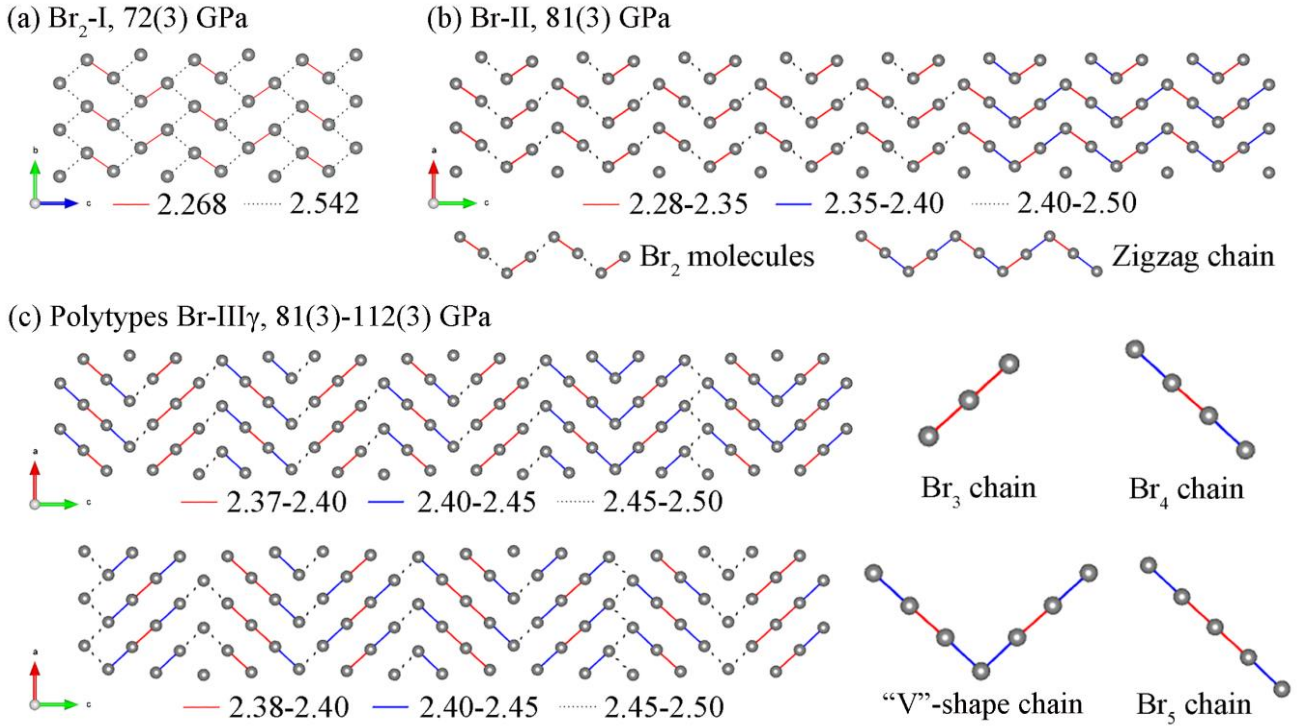

**Fig. 2. Fragments of crystal structures of the bromine allotropes synthesised in this work.** (a) Br$_2$-I (*oC*8) at 72(3) GPa projected to the *bc* plane; the solid red and dashed black lines represent the intramolecular and intermolecular distances (in Å), respectively; (b) Br-II (*Fmmm*(00γ)*s*00; γ = 0.493(3)) at 81(3) GPa projected to the *ac* plane; the bottom figures show the arrangement of bromine atoms in Br-II, which contains both Br$_2$ molecules and Br$_3$ zigzag chains; (c) two examples of the polytypes Br-IIIγ existing in the pressure range of 81(3) to 112(3) GPa: Br-III0.288 at 81(3) GPa and Br-III0.227 at 96(3) GPa (see Table S9) projected to the *ac* plane. Some structural elements (Br$_3$, Br$_4$, Br$_5$, and V-shape (Br$_7$) atomic chains) are also presented. Br atoms are shown as grey balls. Structure presentations of the incommensurate phases in (b) and (c) are approximate (cut from 5×5×20 cells of the basic *oF*4 lattice). For all structures, only one plane at *y* = 0 (*x* = 0 for Br$_2$-I in (a)) is presented for simplicity.

At a given pressure, there are obvious variations of the γ component of the modulation wavevector **q** = γ**c**\* among different crystallites of Br-IIIγ (Table S9). As the difference in γ is small, at first glance it may seem that we are dealing with the same crystalline form of bromine. However, the refined values of γ differ beyond the experimental uncertainty. The analysis of powder XRD data (Fig. S5), collected at the ESRF beamline ID11 with submicron resolution, shows that within a tiny area (0.75 × 0.75 μm$^2$) co-exist numerous domains of Br-IIIγ with distinct values of γ. For example, at one spot the peaks belonging to Br-IIIγ with γ = 0.2266(9), 0.1962(6), and 0.1847(5) appear simultaneously (Fig. S5). There is no evidence of any special spatial distribution of Br-IIIγ crystallites with different γ (Fig. S5 inset), so we suggest that the domains of Br-IIIγ form simultaneously.

Structures of Br-IIIγ with different values of the γ component of the modulation vector can be presented as stacking of distorted close-packed layers of bromine atoms in the *ab* plane along the *c* direction. All layers are identical, as all bromine atoms occupy a single crystallographic position (Fig. S6), and the difference in their stacking is defined by the value of γ. According to [34], "*an element or compound is polytypic if it occurs in*

*several different structural modifications, each of which may be regarded as built up by stacking layers of (nearly) identical structure and composition, and if the modifications differ only in their stacking sequence*". In this sense, one can consider Br-IIIγ compound with the variating value of γ as a series of incommensurate polytypes.

There are striking differences in incommensurate phases Br-II and Br-IIIγ (Fig. 3): while the values of γ determined for all domains of Br-II (as for 81(3) GPa) are the same within the uncertainty, for Br-IIIγ the values of γ vary within ~20% (the uncertainty for γ is only ~1%) at any given pressure point (Fig. 3). The analysis of XRD data from all domains of reasonably good quality (giving $R_{int}$ < 10%) does not reveal any remarkable distribution of the number of crystallites with different values of γ (Fig. 3b).

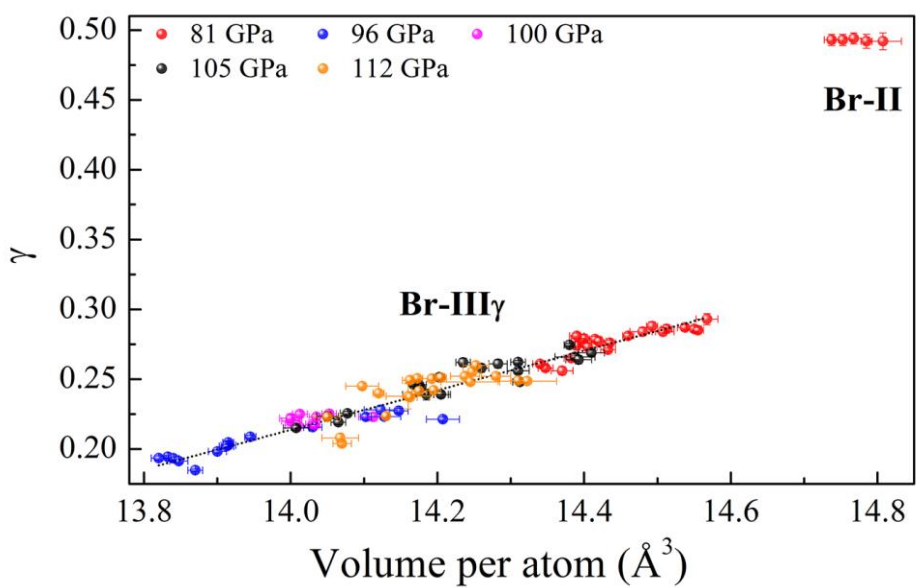

(a)

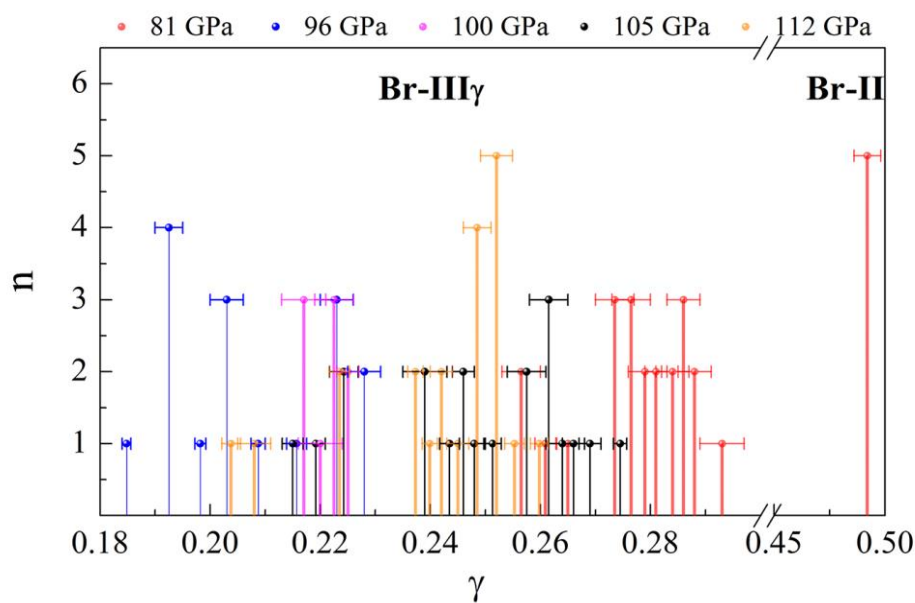

(b)

**Fig. 3. Variations of the γ component of the modulation vector in the structures of Br-II and Br-IIIγ.** (a) The relationship between the values of γ and the volume per atom; the dotted black line is a guide for the eyes. (b) The frequency of the appearance of crystallites (n) with a particular value of γ. The total number of analyzed crystallites at 81(3), 96(3), 100(3), 105(3), and 112(3) GPa was 27, 16, 9, 20, and 21 respectively. The data related to these pressures are shown in red, blue, magenta, black, and orange, respectively. Only single-crystal domains with sufficient quality ($R_{int}$ < 10% and $R_1$ < 20%) were counted.

Upon compression, the volume of $Br_2$-I decreases monotonically between 45(3) and 73(3) GPa (Fig. S7). The experimental data are described by the 2$^{nd}$ order Birch–Murnaghan equation of state (EoS) with the bulk modulus $K_0$ = 22(3) GPa, and the volume at ambient pressure $V_0$ = 32.10(15) Å$^3$/atom. Extrapolation of the volume of $Br_2$-I to 81 GPa suggests that the transition from $Br_2$-I to Br-II does not associate with any significant change in volume (Fig. S7). All analyzed domains of Br-IIIγ at 81(3) GPa show lower atomic volumes than that of Br-II at the same pressure (Fig. S7). We also noticed that the variation (~ 1.6%, 2.7%, 0.8%, 2.8%, and 1.9% at 81, 96, 100, 105, and 112 GPa, respectively) in atomic volumes of different domains of Br-IIIγ is significantly larger than that for Br-II or $Br_2$-I domains at any pressure point (below 0.6%). Moreover, for Br-IIIγ between 81(3) and 112(3) GPa there is no obvious trend in the volume change with pressure (Fig. S7). This observation suggests that incommensurate polytypes of Br-IIIγ with different γ have unique EOSes.

In the whole studied pressure range, for Br-IIIγ there is an obvious linear correction between the volume and the value of γ (Fig. 3a). Considering that the changes in the volume of Br-IIIγ come almost solely due to the changes in the *b* parameter (Fig. S8a), one can notice the decrease of *b* with the decrease of γ (Fig. S8b). Structurally, for Br-IIIγ these mean that at a given pressure the decrease in the shortest distance within the chains (in *ac* plane; D3) of bromine atoms is associated with an increase of the distances between layers of chains (*b* direction; D1 and D2) (Fig. 4a; see Fig. 4b for an illustration of D1, D2, and D3). Remarkably, at the given pressure points, average distances of bromine atoms within the first coordination sphere remain almost the same for a wide range of γ (or volume) (Fig. 4a inset). This observation reminds of the situation with polytypes based on close-packed structures (in metals, for example) – the interatomic distances are the same in the first coordination sphere, and only change in the second one.

The Br-IIIγ phase is structurally similar to that of the recently reported iodine I-V [5,16]. Like in Br-IIIγ, in the pressure range of the existence of the I-V phase (~24 – ~30 GPa) γ decreases with decreasing volume (Fig. S9) [5,16]. We have performed our own experiment on iodine, produced by the decomposition of $CI_4$ upon laser heating at 25.1(5) GPa. Structural data (Table S10) closely reproduce the results of Bykova *et al.* [16] for I-V at similar pressure, however, our analysis of numerous domains revealed a significant variation of γ – from 0.283(2) to 0.300(2) – at a single pressure point. The relation between the volume per atom and γ for iodine I-V is similar to that observed for Br-IIIγ (Fig. S9). Thus, there is a clear indication that incommensurate polytypes are present in I-V phase.

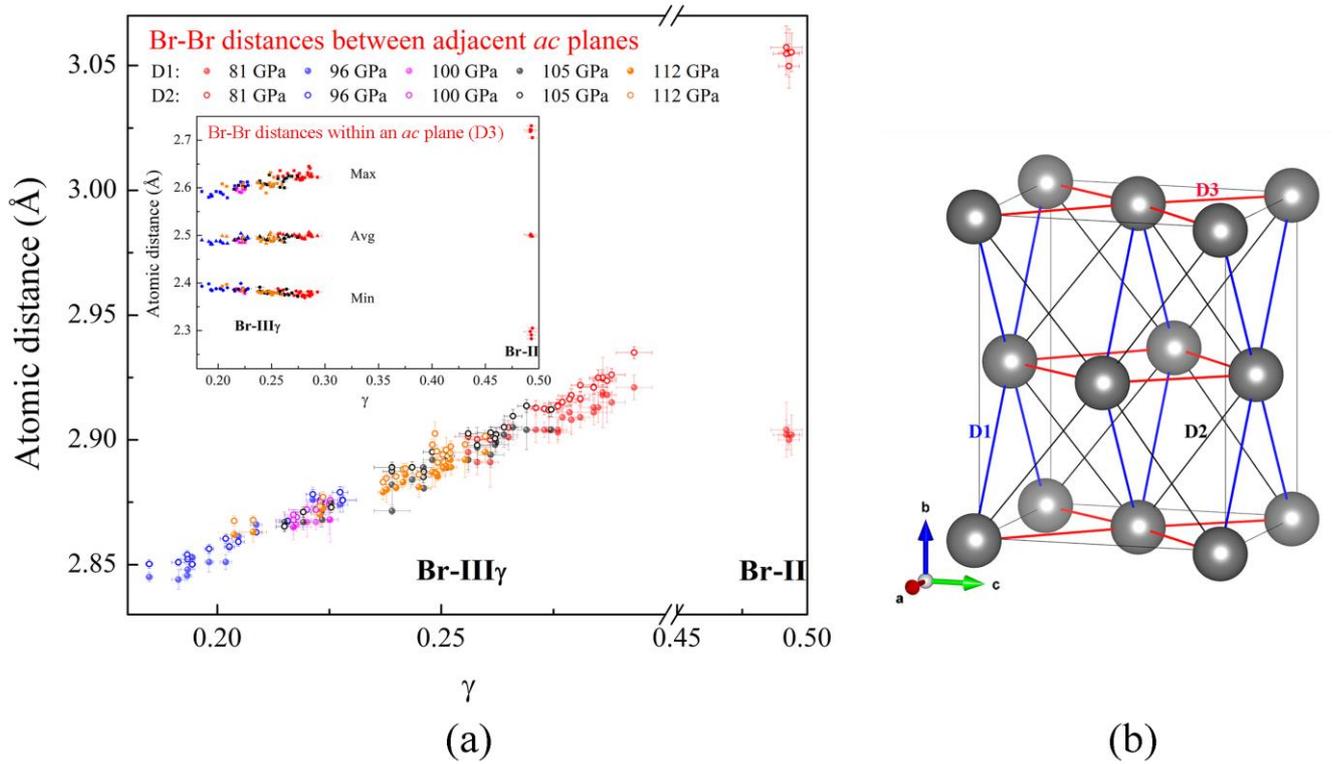

**Fig. 4. Variation of interatomic distances in Br-II and Br-IIIγ as a function of the γ components of the modulation vector.** (a) Relations between the Br-Br interatomic distances (D1, D2, and D3) and γ at different pressures; (b) notations of D1, D2, and D3 in the *i-oF*4 unit cell of bromine. D1 and D2 are the Br-Br distances between the adjacent *ac* planes, and D3 is the distance within the *ac* plane. The four nearest Br-Br distances (D3) in incommensurate phases are continuously distributed between the maximum (Max) and minimum (Min), and their average values are plotted as Avg (inset in (a)).

In summary, we found that bromine preserves its molecular $oC8$ structure (Br$_2$-I) at 72(3) GPa; no sign of other molecular phases was observed. Above 81(3) GPa, we found that bromine forms an incommensurately modulated structure $Fmmm(00\gamma)s00$ (Br-II; $\gamma = 0.493(3)$) and, depending on pressure, several incommensurate polytypes of Br-IIIγ similar to Br-II, but with different values of the γ components (ranging in the interval of ~0.18 to ~0.3). For I-V at 25.1(3) GPa, γ varies from ~0.28 to ~0.30. Thus, Br-IIIγ and I-V are examples of polytypes with incommensurately modulated sequence of layers. The diversity and coexistence of these polytypes suggest that incommensurate polytypism is a common phenomenon for simple halogen systems undergoing transformations from molecular to non-molecular crystal states.

**Acknowledgements**

The authors acknowledge the Deutsches Elektronen-Synchrotron (DESY, PETRA III) and the European Synchrotron Radiation Facility (ESRF) for the provision of beamtime at the P02.2 and, ID11, ID15b, and ID27 beamlines, respectively. Y.Y. acknowledges the financial support provided by the China Scholarship Council (CSC) during her visit to the University of Bayreuth. N.D. and L.D. thank the Deutsche Forschungsgemeinschaft (DFG; projects DU 945–15-1, DU 393–9/2, DU 393–13/1) for financial support. N.D. also thanks the Swedish Government Strategic Research Area in Materials Science on Functional Materials at Linköping University (Faculty Grant SFO-Mat-LiU No. 2009 00971). M.B. acknowledges the support of Deutsche Forschungsgemeinschaft (DFG Emmy-Noether project BY112/2-1). D.L. thanks the UKRI Future Leaders Fellowship (MR/V025724/1) for financial support. For the purpose of open access, the author has applied a Creative Commons Attribution (CC BY) licence to any Author Accepted Manuscript version arising from this submission.

# Supplementary Information

# Polytypism of Incommensurately Modulated Structures of Crystalline Bromine upon Molecular Dissociation under High Pressure

Yuqing Yin[1,2*], Andrey Aslandukov[1,3], Maxim Bykov[4], Dominique Laniel[5], Alena Aslandukova[3], Anna Pakhomova[6], Timofey Fedotenko[7], Wenju Zhou[1], Fariia Iasmin Akbar[1,3], Michael Hanfland[6], Konstantin Glazyrin[7], Carlotta Giacobbe[6], Eleanor Lawrence Bright[6], Gaston Garbarino[6], Zhitai Jia[2], Natalia Dubrovinskaia[1,8], Leonid Dubrovinsky[3*]

**Affiliations:**
[1]Material Physics and Technology at Extreme Conditions, Laboratory of Crystallography, University of Bayreuth, 95440 Bayreuth, Germany
[2]State Key Laboratory of Crystal Materials, Shandong University, Jinan 250100, China
[3]Bayerisches Geoinstitut, University of Bayreuth, 95440 Bayreuth, Germany
[4]Institute of Inorganic Chemistry, University of Cologne, Greinstrasse 6, 50939 Cologne, Germany
[5]Centre for Science at Extreme Conditions and School of Physics and Astronomy, University of Edinburgh, EH9 3FD Edinburgh, United Kingdom
[6]European Synchrotron Radiation Facility, B.P.220, F-38043 Grenoble Cedex, France
[7]Photon Science, Deutsches Elektronen-Synchrotron, Notkestrasse 85, 22607 Hamburg, Germany
[8]Department of Physics, Chemistry and Biology (IFM), Linköping University, SE-581 83, Linköping, Sweden
*Corresponding authors. Email: Yuqing.Yin@uni-bayreuth.de (Y.Y.); Leonid.Dubrovinsky@uni-bayreuth.de (L.D.)

# Supplementary Methods

BX90-type screw-driven diamond anvil cells (DACs) [1] equipped with 120 μm or 250 μm culet diamond anvils were used (Table S1). A piece of $CBr_4$ was loaded in one DAC for the main experiments (DAC #1). A thin (2-5 μm) NaBr plate was loaded together with a piece of $CBr_4$ (DAC #2) as supplementary experiments in order to test the reproducibility of the bromine's polytypism. DAC #3 was loaded with Mg + $CBr_4$, which provided a pressure-volume point for $Br_2$-I. DAC #4 (NaI + $CI_4$) was prepared to test the polytypism behavior of iodine I-V. Rhenium was used as the gasket material. The *in situ* pressure was measured using the first-order Raman mode of the stressed diamond anvils [2] until 81 GPa. At pressures above 81 GPa, the Raman mode from the synthesized diamonds inside the DAC [3] (from the decomposition of $CBr_4$) starts to coincide with the Raman mode of the stressed diamond anvils, resulting in the insufficient quality of the Raman signal. Thus, the *in situ* pressure was measured according to the equation of state of rhenium [4] above 81 GPa. Double-sided sample laser-heating (up to ~2600 K in all cases) was performed at our home laboratory at the Bayerisches Geoinstitut [5] with $CBr_4$ employed as the laser light absorber. Detailed information on pressure can be found in Table S1 and Table S9.

Synchrotron X-ray diffraction measurements of the compressed samples were performed at ID11 (λ = 0.2846 Å, beam size ~0.75 × 0.75 μm$^2$), ID15b (λ = 0.4100 Å, beam size ~ 1.5 × 1.5 μm$^2$), and ID27 (λ = 0.3738 Å, beam size ~ 1.5 × 1.5 μm$^2$) of the EBS-ESRF and the P02.2 beamline (λ = 0.2910 Å, beam size ~2.0 × 2.0 μm$^2$) of PETRA III. In order to determine the sample position for single-crystal X-ray diffraction data acquisition, a full X-ray diffraction mapping of the pressure chamber was performed. The sample positions displaying the greatest number of single-crystal reflections belonging to the phases of interest were chosen, and step-scans of 0.5° from −36° to +36° ω were performed. The CrysAlis$^{Pro}$ software [6] was utilized for the single-crystal data analysis. To calibrate the instrumental model in the CrysAlis$^{Pro}$ software, *i.e.* the sample-to-detector distance, detector's origin, offsets of the goniometer angles, and rotation of both the X-ray beam and detector around the instrument axis, we used a single crystal of orthoenstatite [$(Mg_{1.93}Fe_{0.06})(Si_{1.93}Al_{0.06})O_6$, *Pbca* space group, a = 8.8117(2) Å, b = 5.1832(10) Å, and c = 18.2391(3) Å]. The DAFi program [7] was used for the search of reflections' groups belonging to individual single-crystal domains. The crystal structures were then solved and refined using the OLEX2 [8] and JANA2006 software [9]. The crystal structure visualisation was made with the VESTA software [10]. The equation of state was obtained by fitting the pressure-volume dependence data using the EoSFit7-GUI [11]. The crystallographic information of the bromine allotropes synthesized in this work is available in Tables S2-S8.

## Supplementary Tables

**Table S1. Summary of the performed high-pressure high-temperature experimental syntheses in laser-heated diamond anvil cells.**

| DAC number | Culet diameter (μm) | Starting materials | Pressure points (GPa, ±3) | Bromine phases observed |
|---|---|---|---|---|
| 1 | 120 | $CBr_4$ | 72 | $Br_2$-I |
|  |  |  | 81 | Br-II and Br-III$\gamma$ |
|  |  |  | 96, 105, 112 | Br-III$\gamma$ |
| 2 | 120 | NaBr + $CBr_4$* | 46, 70, 73 | $Br_2$-I |
|  |  |  | 100 | Br-III$\gamma$ |
| 3 | 250 | Mg + $CBr_4$* | 45 | $Br_2$-I |
| DAC number | Culet diameter (μm) | Starting materials | Pressure points (GPa, ±0.5) | Iodine phases observed |
| 4 | 250 | NaI + $CI_4$* | 25.1 | I-V |

* The reagents marked with the star were put into the DACs in excess.

**Table S2. Crystal structure, data collection and refinement details of Br$_2$-I ($oC$8) at 72(3) GPa.**

| Crystal data | |
|---|---|
| Chemical formula | Br |
| $M_r$ | 79.91 |
| Crystal system, space group | Orthorhombic, $Cmce$ |
| Temperature (K) | 293 |
| Pressure (GPa) | 72(3) |
| $a, b, c$ (Å) | 4.867(4), 3.3083(8), 7.5901(14) |
| $V$ (Å$^3$) | 122.22 (11) |
| $Z$ | 8 |
| Radiation type | Synchrotron, $\lambda$ = 0.291 Å |
| $\mu$ (mm$^{-1}$) | 5.85 |
| Crystal size (mm) | 0.001 × 0.001 × 0.001 |
| **Data collection** | |
| Diffractometer | LH@P02.2 |
| No. of measured, independent and observed [$I > 2\sigma(I)$] reflections | 309, 128, 126 |
| $R_{int}$ | 0.009 |
| $(\sin \theta/\lambda)_{max}$ (Å$^{-1}$) | 1.015 |
| **Refinement** | |
| $R[F^2 > 2\sigma(F^2)]$, $wR(F^2)$, $S$ | 0.033, 0.079, 1.41 |
| No. of reflections | 128 |
| No. of parameters | 8 |
| $\Delta\rho_{max}$, $\Delta\rho_{min}$ (e Å$^{-3}$) | 1.48, -1.73 |
| **Crystal Structure** | |
| Wyckoff Site | Br1: 8$f$ |
| Fractional atomic coordinates (x y z) | Br1: 1/2 0.3050(2) 0.62284(7) |
| $U_{iso}$ (Å$^2$) | Br1: 0.0054(3) |

**Table S3. Crystal structure, data collection and refinement details of incommensurate Br-II at 81(3) GPa.** Left column shows the refinement using only the 1st order satellite reflections. Right column – using the reflections of both the 1st and 2nd order.

| Crystal data | | |
|---|---|---|
| Chemical formula | Br | Br |
| $M_r$ | 79.91 | 79.91 |
| Crystal system, space group | Orthorhombic, $Fmmm(00\gamma)s00$ | Orthorhombic, $Fmmm(00\gamma)s00$ |
| Temperature (K) | 293 | 293 |
| Pressure (GPa) | 81(3) | 81(3) |
| Wave vectors | **q** = 0.493(3)**c**$^*$ | **q** = 0.494(3)**c**$^*$ |
| $a, b, c$ (Å) | 3.2767(14), 4.7924(19), 3.7618(16) | 3.272(2), 4.789(3), 3.759(3) |
| $V$ (Å$^3$) | 59.07(4) | 58.90(7) |
| $Z$ | 4 | 4 |
| Radiation type | Synchrotron, $\lambda$ = 0.41 Å | Synchrotron, $\lambda$ = 0.41 Å |
| $\mu$ (mm$^{-1}$) | 15.71 | 15.76 |
| Crystal size (mm) | 0.001 × 0.001 × 0.001 | 0.001 × 0.001 × 0.001 |
| **Data collection** | | |
| Diffractometer | ESRF ID15b, EIGER2 X 9M CdTe detector | ESRF ID15b, EIGER2 X 9M CdTe detector |
| No. of main reflections: measured, independent and observed [$I > 3\sigma(I)$] | 33, 20, 19 | 29, 20, 19 |
| No. of 1st-order satellite reflections: measured, independent and observed [$I > 3\sigma(I)$] | 51, 36, 36 | 46, 33, 33 |
| No. of 2nd-order satellite reflections: measured, independent and observed [$I > 3\sigma(I)$] | - | 51, 33, 20 |
| $R_{int}$ | 0.010 | 0.028 |
| $(\sin \theta/\lambda)_{max}$ (Å$^{-1}$) | 0.879 | 0.880 |
| **Refinement** | | |
| $R(obs)_{main + satellites}$ / $wR(all)_{main + satellites}$ | 0.060 / 0.085 | 0.105 / 0.145 |
| $R(obs)_{main}$ / $wR(all)_{main}$ | 0.051 / 0.069 | 0.072 / 0.093 |
| $R(obs)_{1st\ order\ satellites}$ / $wR(all)_{1st\ order\ satellites}$ | 0.071 / 0.093 | 0.140 / 0.167 |
| $R(obs)_{2nd\ order\ satellites}$ / $wR(all)_{2nd\ order\ satellites}$ | - | 0.150 / 0.159 |
| No. of reflections | 56 | 86 |
| No. of parameters | 5 | 6 |
| $\Delta\rho_{max}, \Delta\rho_{min}$ (e Å$^{-3}$) | 4.33, -4.15 | 5.95, -9.82 |
| **Crystal Structure** | | |

| Br (x y z) | (0 0 0) | (0 0 0) |
|---|---|---|
| $A_1(x)$ | 0.0742(11) | -0.0684(12) |
| $A_2(y)$ | - | -0.0032(10) |

**Table S4.** Crystal structure, data collection and refinement details of incommensurate Br-IIIγ at 81(3) GPa.

| Crystal data | | | |
|---|---|---|---|
| Chemical formula | Br | Br | Br |
| $M_r$ | 79.91 | 79.91 | 79.91 |
| Crystal system, space group | Orthorhombic, $Fmmm(00\gamma)s00$ | Orthorhombic, $Fmmm(00\gamma)s00$ | Orthorhombic, $Fmmm(00\gamma)s00$ |
| Temperature (K) | 293 | 293 | 293 |
| Pressure (GPa) | 81(3) | 81(3) | 81(3) |
| Wave vectors | **q** = 0.288(2)**c**\* | **q** = 0.2708(12)**c**\* | **q** = 0.2577(18)**c**\* |
| a, b, c (Å) | 3.5179(11), 4.6502(12), 3.5441(10) | 3.5315(7), 4.611(2), 3.5390(5) | 3.5297(7), 4.582(4), 3.5529(17) |
| V (Å$^3$) | 57.98(3) | 57.63(3) | 57.46(6) |
| Z | 4 | 4 | 4 |
| Radiation type | Synchrotron, λ = 0.41 Å | Synchrotron, λ = 0.41 Å | Synchrotron, λ = 0.41 Å |
| μ (mm$^{-1}$) | 16.01 | 16.10 | 16.15 |
| Crystal size (mm) | 0.001 × 0.001 × 0.001 | 0.001 × 0.001 × 0.001 | 0.001 × 0.001 × 0.001 |
| **Data collection** | | | |
| Diffractometer | ESRF ID15b, EIGER2 X 9M CdTe detector | ESRF ID15b, EIGER2 X 9M CdTe detector | ESRF ID15b, EIGER2 X 9M CdTe detector |
| No. of main reflections: measured, independent and observed [$I > 3\sigma(I)$] | 41, 26, 26 | 37, 22, 22 | 34, 20, 20 |
| No. of 1$^{st}$-order satellite reflections: measured, independent and observed [$I > 3\sigma(I)$] | 58, 36, 36 | 71, 39, 39 | 60, 35, 35 |
| No. of 2$^{nd}$-order satellite reflections: measured, independent and observed [$I > 3\sigma(I)$] | 68, 39, 13 | 68, 41, 30 | 73, 40, 19 |
| $R_{int}$ | 0.009 | 0.011 | 0.025 |
| $(\sin \theta/\lambda)_{max}$ (Å$^{-1}$) | 0.864 | 0.864 | 0.851 |
| **Refinement** | | | |
| $R(obs)_{main\ +\ satellites}$ / $wR(all)_{main\ +\ satellites}$ | 0.049 / 0.065 | 0.078 / 0.097 | 0.075 / 0.084 |
| $R(obs)_{main}$ / $wR(all)_{main}$ | 0.048 / 0.069 | 0.071 / 0.079 | 0.071 / 0.082 |
| $R(obs)_{\text{1st order satellites}}$ / $wR(all)_{\text{1st order satellites}}$ | 0.044 / 0.056 | 0.081 / 0.103 | 0.060 / 0.077 |
| $R(obs)_{\text{2nd order satellites}}$ / $wR(all)_{\text{2nd order satellites}}$ | 0.142 / 0.156 | 0.094 / 0.108 | 0.165 / 0.165 |
| No. of reflections | 101 | 102 | 95 |
| No. of parameters | 6 | 6 | 6 |
| $\Delta\rho_{max}, \Delta\rho_{min}$ (e Å$^{-3}$) | 2.70, -2.65 | 2.86, -3.18 | 4.12, -3.38 |
| **Crystal Structure** | | | |
| Br (*x y z*) | (0 0 0) | (0 0 0) | (0 0 0) |
| $A_1(x)$ | 0.0569(6) | -0.0607(6) | -0.0631(8) |
| $A_2(y)$ | -0.0002(9) | 0.0004(6) | -0.0022(16) |

**Table S5. Crystal structure, data collection and refinement details of incommensurate Br-IIIγ at 96(3) GPa.**

| Crystal data | | | |
|---|---|---|---|
| Chemical formula | Br | Br | Br |
| $M_r$ | 79.91 | 79.91 | 79.91 |
| Crystal system, space group | Orthorhombic, $Fmmm(00\gamma)s00$ | Orthorhombic, $Fmmm(00\gamma)s00$ | Orthorhombic, $Fmmm(00\gamma)s00$ |
| Temperature (K) | 293 | 293 | 293 |
| Pressure (GPa) | 96(3) | 96(3) | 96(3) |
| Wave vectors | **q** = 0.2266(9)**c*** | **q** = 0.1962(6)**c*** | **q** = 0.1847(5)**c*** |
| $a, b, c$ (Å) | 3.5234(8), 4.5340(9), 3.5324(7) | 3.5164(5), 4.4936(5), 3.5002(14) | 3.5153(12), 4.474(2), 3.527(2) |
| $V$ (Å$^3$) | 56.43(2) | 55.31(2) | 55.47(4) |
| $Z$ | 4 | 4 | 4 |
| Radiation type | Synchrotron, $\lambda$ = 0.2846 Å | Synchrotron, $\lambda$ = 0.2846 Å | Synchrotron, $\lambda$ = 0.2846 Å |
| $\mu$ (mm$^{-1}$) | 5.97 | 6.09 | 6.07 |
| Crystal size (mm) | 0.001 × 0.001 × 0.001 | 0.001 × 0.001 × 0.001 | 0.001 × 0.001 × 0.001 |
| **Data collection** | | | |
| Diffractometer | ESRF ID11, Dectris Eiger2 X CdTe 4M | ESRF ID11, Dectris Eiger2 X CdTe 4M | ESRF ID11, Dectris Eiger2 X CdTe 4M |
| No. of main reflections: measured, independent and observed [$I > 3\sigma(I)$] | 92, 52, 51 | 89, 33, 25 | 97, 46, 39 |
| No. of 1$^{st}$-order satellite reflections: measured, independent and observed [$I > 3\sigma(I)$] | 166, 83, 83 | 171, 50, 50 | 178, 79, 78 |
| No. of 2$^{nd}$-order satellite reflections: measured, independent and observed [$I > 3\sigma(I)$] | 179, 96, 45 | 182, 55, 43 | 186, 87, 61 |
| $R_{int}$ | 0.008 | 0.028 | 0.016 |
| $(\sin \theta/\lambda)_{max}$ (Å$^{-1}$) | 1.139 | 1.136 | 1.124 |
| **Refinement** | | | |
| $R(obs)_{main\ +\ satellites}$ / $wR(all)_{main\ +\ satellites}$ | 0.039 / 0.045 | 0.032 / 0.035 | 0.038 / 0.044 |
| $R(obs)_{main}$ / $wR(all)_{main}$ | 0.033 / 0.042 | 0.023 / 0.029 | 0.033 / 0.047 |
| $R(obs)_{\text{1st order satellites}}$ / $wR(all)_{\text{1st order satellites}}$ | 0.033 / 0.038 | 0.028 / 0.031 | 0.036 / 0.039 |
| $R(obs)_{\text{2nd order satellites}}$ / $wR(all)_{\text{2nd order satellites}}$ | 0.109 / 0.124 | 0.059 / 0.063 | 0.054 / 0.052 |
| No. of reflections | 231 | 138 | 212 |
| No. of parameters | 6 | 6 | 6 |
| $\Delta\rho_{max}, \Delta\rho_{min}$ (e Å$^{-3}$) | 4.89, -2.26 | 2.60, -3.29 | 3.29, -3.24 |
| **Crystal Structure** | | | |
| Br ($x\ y\ z$) | (0 0 0) | (0 0 0) | (0 0 0) |
| $A_1(x)$ | 0.0625(3) | 0.0652(2) | 0.0697(2) |
| $A_2(y)$ | -0.0002(3) | 0.0005(7) | 0.0000(4) |

**Table S6.** Crystal structure, data collection and refinement details of incommensurate Br-III$\gamma$ at 100(3) GPa.

| Crystal data | |
|---|---|
| Chemical formula | Br |
| $M_r$ | 79.91 |
| Crystal system, space group | Orthorhombic, $Fmmm(00\gamma)s00$ |
| Temperature (K) | 293 |
| Pressure (GPa) | 100(3) |
| Wave vectors | **q** = 0.223(2)**c**\* |
| $a, b, c$ (Å) | 3.5152(11), 4.5371(17), 3.5202(17) |
| $V$ (Å$^3$) | 56.14(4) |
| $Z$ | 4 |
| Radiation type | Synchrotron, $\lambda$ = 0.3738 Å |
| $\mu$ (mm$^{-1}$) | 12.82 |
| Crystal size (mm) | 0.001 × 0.001 × 0.001 |
| **Data collection** | |
| Diffractometer | ESRF ID27, EIGER2 X CdTe 9M detector |
| No. of main reflections: measured, independent and observed [$I > 3\sigma(I)$] | 35, 18, 18 |
| No. of 1$^{st}$-order satellite reflections: measured, independent and observed [$I > 3\sigma(I)$] | 68, 38, 36 |
| No. of 2$^{nd}$-order satellite reflections: measured, independent and observed [$I > 3\sigma(I)$] | - |
| $R_{int}$ | 0.022 |
| $(\sin\theta/\lambda)_{max}$ (Å$^{-1}$) | 0.889 |
| **Refinement** | |
| $R(obs)_{main + satellites}$ / $wR(all)_{main + satellites}$ | 0.061 / 0.072 |
| $R(obs)_{main}$ / $wR(all)_{main}$ | 0.053 / 0.072 |
| $R(obs)_{1st\ order\ satellites}$ / $wR(all)_{1st\ order\ satellites}$ | 0.068 / 0.072 |
| $R(obs)_{2nd\ order\ satellites}$ / $wR(all)_{2nd\ order\ satellites}$ | - |
| No. of reflections | 56 |
| No. of parameters | 5 |
| $\Delta\rho_{max}, \Delta\rho_{min}$ (e Å$^{-3}$) | 3.51, -3.60 |
| **Crystal Structure** | |
| Br ($x\ y\ z$) | (0 0 0) |
| $A_1(x)$ | -0.0629(10) |
| $A_2(y)$ | - |

**Table S7. Crystal structure, data collection and refinement details of incommensurate Br-IIIγ at 105(3) GPa.**

| Crystal data | | | | |
|---|---|---|---|---|
| Chemical formula | Br | Br | Br | Br |
| $M_r$ | 79.91 | 79.91 | 79.91 | 79.91 |
| Crystal system, space group | Orthorhombic, $Fmmm(00\gamma)s00$ | Orthorhombic, $Fmmm(00\gamma)s00$ | Orthorhombic, $Fmmm(00\gamma)s00$ | Orthorhombic, $Fmmm(00\gamma)s00$ |
| Temperature (K) | 293 | 293 | 293 | 293 |
| Pressure (GPa) | 105(3) | 105(3) | 105(3) | 105(3) |
| Wave vectors | **q** = 0.2745(12)**c**\* | **q** = 0.266(2)**c**\* | **q** = 0.2517(16)**c**\* | **q** = 0.2192(17)**c**\* |
| $a, b, c$ (Å) | 3.5195(9), 4.6201(9), 3.5372(17) | 3.528(2), 4.616(2), 3.534(6) | 3.5164(10), 4.5850(15), 3.5239(14) | 3.5244(13), 4.5298(18), 3.5265(13) |
| $V$ (Å$^3$) | 57.52(3) | 57.55(11) | 56.81(3) | 56.30(4) |
| Z | 4 | 4 | 4 | 4 |
| Radiation type | Synchrotron, λ = 0.291 Å | Synchrotron, λ = 0.291 Å | Synchrotron, λ = 0.291 Å | Synchrotron, λ = 0.291 Å |
| μ (mm$^{-1}$) | 6.18 | 6.18 | 6.26 | 6.32 |
| Crystal size (mm) | 0.001 × 0.001 × 0.001 | 0.001 × 0.001 × 0.001 | 0.001 × 0.001 × 0.001 | 0.001 × 0.001 × 0.001 |
| **Data collection** | | | | |
| Diffractometer | LH@P02.2 | LH@P02.2 | LH@P02.2 | LH@P02.2 |
| No. of main reflections: measured, independent and observed [$I > 3\sigma(I)$] | 39, 26, 23 | 45, 29, 23 | 83, 38, 33 | 42, 26, 25 |
| No. of 1$^{st}$-order satellite reflections: measured, independent and observed [$I > 3\sigma(I)$] | 74, 41, 41 | 89, 51, 49 | 127, 49, 49 | 68, 41, 40 |
| No. of 2$^{nd}$-order satellite reflections: measured, independent and observed [$I > 3\sigma(I)$] | 84, 44, 30 | - | 139, 62, 25 | 86, 50, 20 |
| $R_{int}$ | 0.015 | 0.023 | 0.025 | 0.039 |
| $(\sin\theta/\lambda)_{max}$ (Å$^{-1}$) | 1.006 | 1.036 | 1.041 | 0.833 |
| **Refinement** | | | | |
| $R(obs)_{main + satellites}$ / $wR(all)_{main + satellites}$ | 0.054 / 0.069 | 0.054 / 0.061 | 0.050 / 0.062 | 0.070 / 0.085 |
| $R(obs)_{main}$ / $wR(all)_{main}$ | 0.061 / 0.079 | 0.051 / 0.062 | 0.045 / 0.053 | 0.065 / 0.088 |
| $R(obs)_{1st\ order\ satellites}$ / $wR(all)_{1st\ order\ satellites}$ | 0.047 / 0.063 | 0.056 / 0.061 | 0.050 / 0.054 | 0.062 / 0.073 |

| $R(obs)_{\text{2nd order satellites}}$ / $wR(all)_{\text{2nd order satellites}}$ | 0.054 / 0.060 | - | 0.080 / 0.177 | 0.133 / 0.137 |
|---|---|---|---|---|
| No. of reflections | 111 | 80 | 149 | 117 |
| No. of parameters | 6 | 5 | 6 | 6 |
| $\Delta\rho_{max}$, $\Delta\rho_{min}$ (e Å$^{-3}$) | 2.40, -2.86 | 3.38, -2.35 | 2.86, -3.05 | 4.59, -4.11 |
| **Crystal Structure** | | | | |
| Br ($x\ y\ z$) | (0 0 0) | (0 0 0) | (0 0 0) | (0 0 0) |
| $A_1(x)$ | -0.0606(4) | -0.0620(7) | 0.0576(4) | 0.0656(7) |
| $A_2(y)$ | -0.0007(13) | - | 0.0004(9) | -0.0008(15) |

Table S8. Crystal structure, data collection and refinement details of incommensurate Br-IIIγ at 112(3) GPa.

| Crystal data | | | |
|---|---|---|---|
| Chemical formula | Br | Br | Br |
| $M_r$ | 79.91 | 79.91 | 79.91 |
| Crystal system, space group | Orthorhombic, $Fmmm(00\gamma)s00$ | Orthorhombic, $Fmmm(00\gamma)s00$ | Orthorhombic, $Fmmm(00\gamma)s00$ |
| Temperature (K) | 293 | 293 | 293 |
| Pressure (GPa) | 112(3) | 112(3) | 112(3) |
| Wave vectors | **q** = 0.2503(12)**c*** | **q** = 0.2370(12)**c*** | **q** = 0.2038(17)**c*** |
| $a, b, c$ (Å) | 3.5137(8), 4.5896(10), 3.520(5) | 3.5249(14), 4.554(3), 3.5291(6) | 3.5278(6), 4.5076(7), 3.539(3) |
| $V$ (Å$^3$) | 56.77(8) | 56.65(4) | 56.28(5) |
| $Z$ | 4 | 4 | 4 |
| Radiation type | Synchrotron, $\lambda$ = 0.2846 Å | Synchrotron, $\lambda$ = 0.2846 Å | Synchrotron, $\lambda$ = 0.2846 Å |
| $\mu$ (mm$^{-1}$) | 5.93 | 5.95 | 5.99 |
| Crystal size (mm) | 0.001 × 0.001 × 0.001 | 0.001 × 0.001 × 0.001 | 0.001 × 0.001 × 0.001 |
| **Data collection** | | | |
| Diffractometer | ESRF ID11, Dectris Eiger2 X CdTe 4M | ESRF ID11, Dectris Eiger2 X CdTe 4M | ESRF ID11, Dectris Eiger2 X CdTe 4M |
| No. of main reflections: measured, independent and observed [$I > 3\sigma(I)$] | 78, 36, 30 | 74, 41, 30 | 60, 23, 21 |
| No. of 1$^{st}$-order satellite reflections: measured, independent and observed [$I > 3\sigma(I)$] | 144, 48, 45 | 145, 77, 73 | 99, 30, 29 |
| No. of 2$^{nd}$-order satellite reflections: measured, independent and observed [$I > 3\sigma(I)$] | 159, 52, 27 | - | - |
| $R_{int}$ | 0.017 | 0.027 | 0.073 |
| $(\sin\theta/\lambda)_{max}$ (Å$^{-1}$) | 1.063 | 1.128 | 0.802 |
| **Refinement** | | | |
| $R(obs)_{main\ +\ satellites}$ / $wR(all)_{main\ +\ satellites}$ | 0.033 / 0.034 | 0.044 / 0.046 | 0.039 / 0.047 |
| $R(obs)_{main}$ / $wR(all)_{main}$ | 0.024 / 0.028 | 0.035 / 0.040 | 0.038 / 0.048 |
| $R(obs)_{\text{1st order satellites}}$ / $wR(all)_{\text{1st order satellites}}$ | 0.034 / 0.030 | 0.050 / 0.050 | 0.039 / 0.047 |
| $R(obs)_{\text{2nd order satellites}}$ / $wR(all)_{\text{2nd order satellites}}$ | 0.076 / 0.114 | - | - |
| No. of reflections | 136 | 118 | 53 |
| No. of parameters | 6 | 5 | 5 |
| $\Delta\rho_{max}, \Delta\rho_{min}$ (e Å$^{-3}$) | 1.78, -1.51 | 2.75, -2.46 | 1.18, -1.34 |
| **Crystal Structure** | | | |
| Br (*x y z*) | (0 0 0) | (0 0 0) | (0 0 0) |
| $A_1(x)$ | 0.0587(2) | 0.0634(4) | 0.0687(9) |
| $A_2(y)$ | 0.0003(11) | - | - |

**Table S9. Bromine allotropes synthesized at high pressures*.**

| Phase | Pressure (GPa, ±3) | Space group | Lattice parameters, *a b c* (Å) | γ | CSD number |
|---|---|---|---|---|---|
| Br$_2$-I | 72 | *Cmce* | 4.867(4) 3.3083(8) 7.5901(14) | - | |
| Br-II | 81 | | 3.2767(14) 4.7924(19) 3.7618(16) | 0.493(3) | 2289479 |
| Br-IIIγ | 81 | *Fmmm*(00γ)*s*00 | 3.515(2) 4.667(3) 3.552(2) | 0.293(4) | |
| | | | 3.5179(11) 4.6502(12) 3.5441(10) | 0.288(2) | 2289480 |
| | | | 3.523(2) 4.636(3) 3.546(3) | 0.284(3) | |
| | | | 3.5163(11) 4.6360(10) 3.531(2) | 0.281(2) | |
| | | | 3.5222(4) 4.6348(13) 3.5321(17) | 0.2786(19) | |
| | | | 3.5216(18) 4.6176(15) 3.5399(10) | 0.274(3) | |
| | | | 3.5315(7) 4.611(2) 3.5390(5) | 0.2708(12) | 2289481 |
| | | | 3.5339(6) 4.601(3) 3.5385(10) | 0.265(2) | |
| | | | 3.5297(7) 4.582(4) 3.5529(17) | 0.2577(18) | 2289482 |
| | 96 | | 3.5234(8) 4.5340(9) 3.5324(7) | 0.2266(9) | 2289483 |
| | | | 3.528(4) 4.530(3) 3.536(3) | 0.2232(17) | |
| | | | 3.527(2) 4.5222(13) 3.5183(19) | 0.2157(18) | |
| | | | 3.5084(9) 4.5251(15) 3.5133(8) | 0.2087(13) | |
| | | | 3.5181(12) 4.5133(8) 3.5054(11) | 0.2047(13) | |
| | | | 3.5055(14) 4.497(2) 3.5300(13) | 0.2019(13) | |
| | | | 3.5164(5) 4.4936(5) 3.5002(14) | 0.1962(6) | 2289484 |
| | | | 3.509(2) 4.477(2) 3.5260(19) | 0.1913(15) | |
| | | | 3.5153(12) 4.474(2) 3.527(2) | 0.1847(5) | 2289485 |
| | 100 | | 3.5105(19) 4.5352(14) 3.5306(7) | 0.2252(17) | |
| | | | 3.5152(11) 4.5371(17) 3.5202(17) | 0.223(2) | 2289486 |
| | | | 3.5174(9) 4.5221(9) 3.5287(11) | 0.217(2) | |
| | 105 | | 3.5195(9) 4.6201(9) 3.5372(17) | 0.2745(12) | 2289487 |
| | | | 3.521(3) 4.6182(17) 3.545(3) | 0.269(2) | |
| | | | 3.528(2) 4.616(2) 3.534(6) | 0.266(2) | 2289488 |
| | | | 3.5141(19) 4.5991(13) 3.5351(18) | 0.261(3) | |
| | | | 3.5182(10) 4.592(2) 3.543(3) | 0.256(2) | |
| | | | 3.5164(10) 4.5850(15) 3.5239(14) | 0.2517(16) | 2289489 |
| | | | 3.5177(9) 4.570(2) 3.5279(10) | 0.2435(18) | |
| | | | 3.525(2) 4.5616(8) 3.5335(8) | 0.239(2) | |
| | | | 3.5285(9) 4.5391(11) 3.516(2) | 0.2254(15) | |
| | | | 3.5244(13) 4.5298(18) 3.5265(13) | 0.2192(17) | 2289490 |
| | | | 3.5260(13) 4.521(2) 3.515(4) | 0.2150(19) | |
| | 112 | | 3.5137(8) 4.5896(10) 3.520(5) | 0.2503(12) | 2289491 |
| | | | 3.507(3) 4.5729(13) 3.516(5) | 0.245(2) | |
| | | | 3.5249(14) 4.554(3) 3.5291(6) | 0.2370(12) | 2289492 |

|  |  |  | 3.5260(9) 4.5356(10) 3.5339(7) | 0.2236(16) |  |
|  |  |  | 3.5278(6) 4.5076(7) 3.539(3) | 0.2038(17) | 2289493 |

*For Br-IIIγ, only selected data are presented.

**Table S10.** Crystal structure, data collection and refinement details of incommensurate I-V at 25.1(5) GPa.

| Crystal data | | |
|---|---|---|
| Chemical formula | I | I |
| $M_r$ | 126.9 | 126.9 |
| Crystal system, space group | Orthorhombic, $Fmmm(00\gamma)s00$ | Orthorhombic, $Fmmm(00\gamma)s00$ |
| Temperature (K) | 293 | 293 |
| Pressure (GPa) | 25.1(5) | 25.1(5) |
| Wave vectors | **q** = 0.2888(18)**c*** | **q** = 0.3004(15)**c*** |
| $a, b, c$ (Å) | 4.2254(13), 5.536(3), 4.249(2) | 4.2221(13), 5.569(6), 4.2550(11) |
| $V$ (Å$^3$) | 99.39(8) | 100.05(12) |
| Z | 4 | 4 |
| Radiation type | Synchrotron, $\lambda$ = 0.41 Å | Synchrotron, $\lambda$ = 0.41 Å |
| $\mu$ (mm$^{-1}$) | 6.99 | 6.95 |
| Crystal size (mm) | 0.001 × 0.001 × 0.001 | 0.001 × 0.001 × 0.001 |
| **Data collection** | | |
| Diffractometer | ESRF ID15b, EIGER2 X 9M CdTe detector | ESRF ID15b, EIGER2 X 9M CdTe detector |
| No. of main reflections: measured, independent and observed [$I > 3\sigma(I)$] | 46, 33, 30 | 58, 34, 31 |
| No. of 1$^{st}$-order satellite reflections: measured, independent and observed [$I > 3\sigma(I)$] | 74, 50, 48 | 100, 58, 55 |
| No. of 2$^{nd}$-order satellite reflections: measured, independent and observed [$I > 3\sigma(I)$] | 95, 58, 27 | 112, 60, 31 |
| $R_{int}$ | 0.012 | 0.035 |
| $(\sin\theta/\lambda)_{max}$ (Å$^{-1}$) | 0.872 | 0.855 |
| **Refinement** | | |
| $R(obs)_{main + satellites}$ / $wR(all)_{main + satellites}$ | 0.044 / 0.048 | 0.105 / 0.117 |
| $R(obs)_{main}$ / $wR(all)_{main}$ | 0.041 / 0.046 | 0.106 / 0.113 |
| $R(obs)_{1st\ order\ satellites}$ / $wR(all)_{1st\ order\ satellites}$ | 0.035 / 0.041 | 0.092 / 0.115 |
| $R(obs)_{2nd\ order\ satellites}$ / $wR(all)_{2nd\ order\ satellites}$ | 0.098 / 0.104 | 0.154 / 0.204 |
| No. of reflections | 141 | 152 |
| No. of parameters | 6 | 6 |

| Δρ$_{max}$, Δρ$_{min}$ (e Å$^{-3}$) | 2.70, -1.97 | 6.68, -5.02 |
|---|---|---|
| **Crystal Structure** | | |
| Br (*x y z*) | (0 0 0) | (0 0 0) |
| $A_1$(*x*) | -0.0599(3) | -0.0587(8) |
| $A_2$(*y*) | -0.0016(7) | -0.0028(15) |

# Supplementary Figures

## Molecular structures

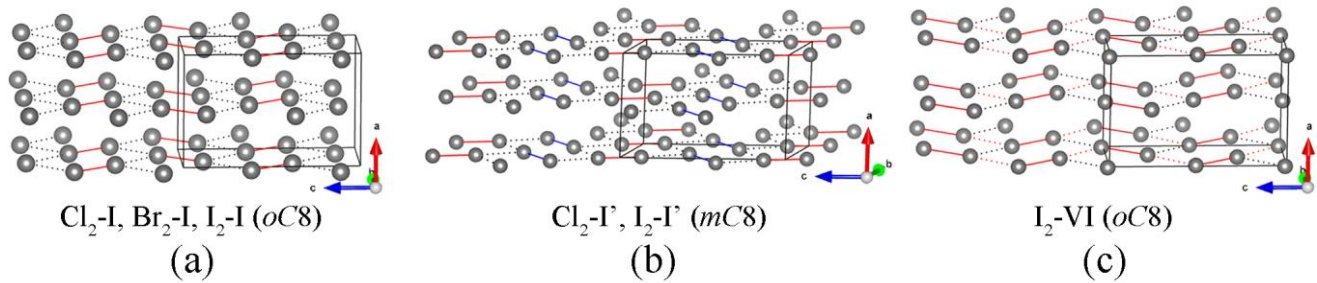

(a) Cl$_2$-I, Br$_2$-I, I$_2$-I (*oC*8)  (b) Cl$_2$-I', I$_2$-I' (*mC*8)  (c) I$_2$-VI (*oC*8)

## Incommensurately modulated structures of Cl and I

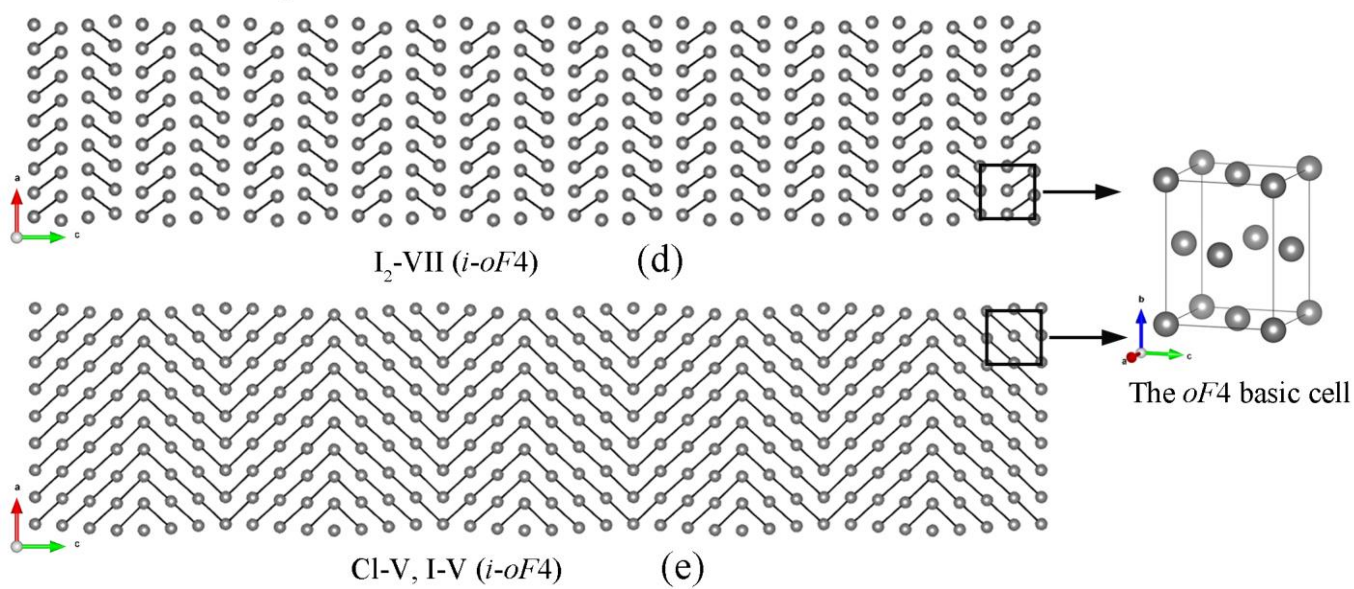

(d) I$_2$-VII (*i-oF*4)

(e) Cl-V, I-V (*i-oF*4)

The *oF*4 basic cell

## Non-molecular structures

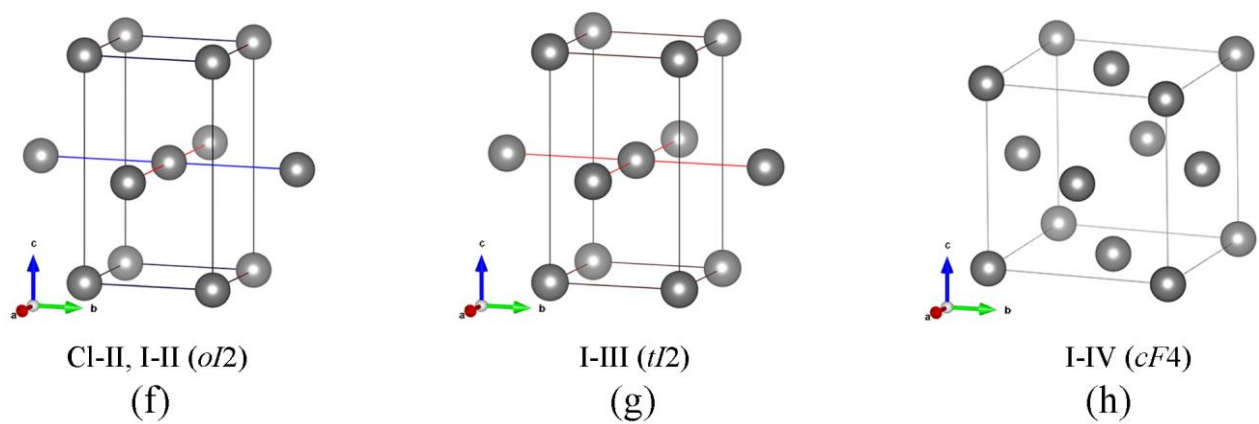

(f) Cl-II, I-II (*oI*2)  (g) I-III (*tI*2)  (h) I-IV (*cF*4)

Fig. S1. Crystal structures of allotropes of I, Br, and Cl (see also the schematic diagram in Fig. 1) [12-19].

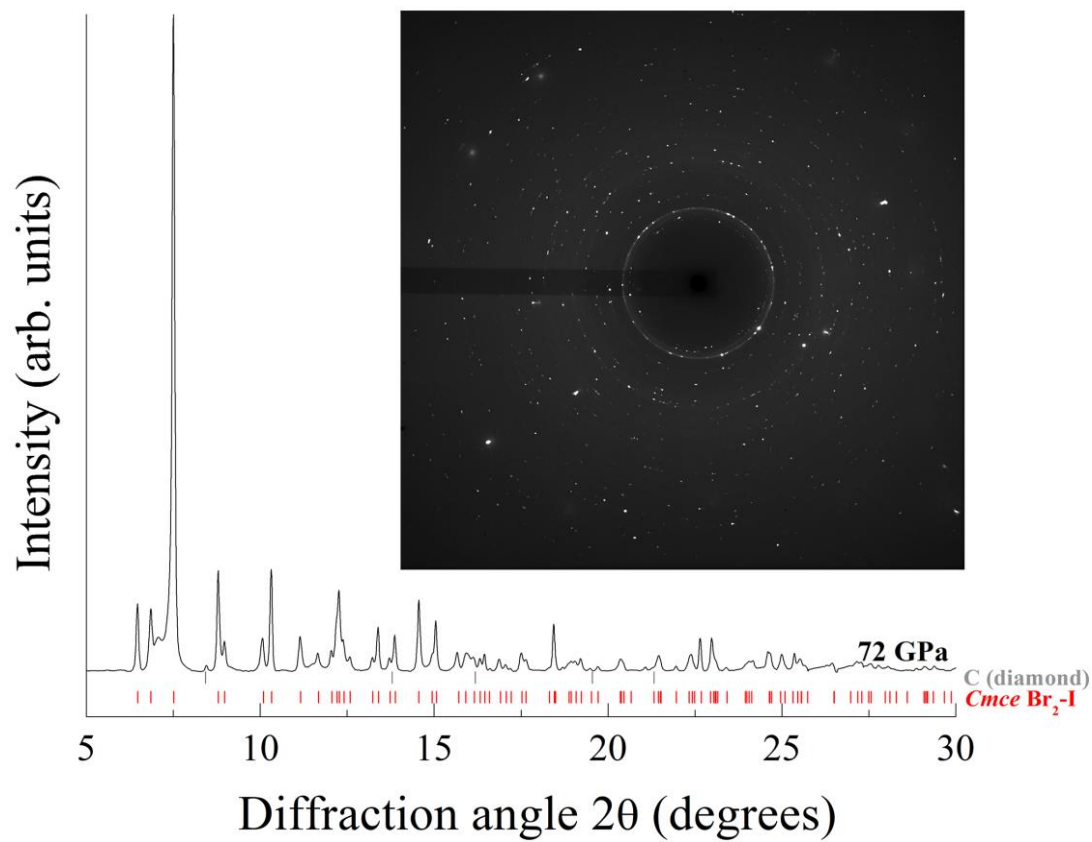

Fig. S2. XRD pattern of the sample in DAC #1 at 72(3) GPa after its laser heating at ~ 2600 K. Black line is the integrated 2D XRD pattern (in inset) of the sample. Red ticks correspond to the calculated powder XRD pattern of the *Cmce* $Br_2$-I (*oC*8) phase ($\lambda = 0.291$ Å), whose structure was determined from SCXRD data. All of the diffraction peaks can be assigned to *Cmce* $Br_2$-I, thus confirming the absence of any other phases in the sample.

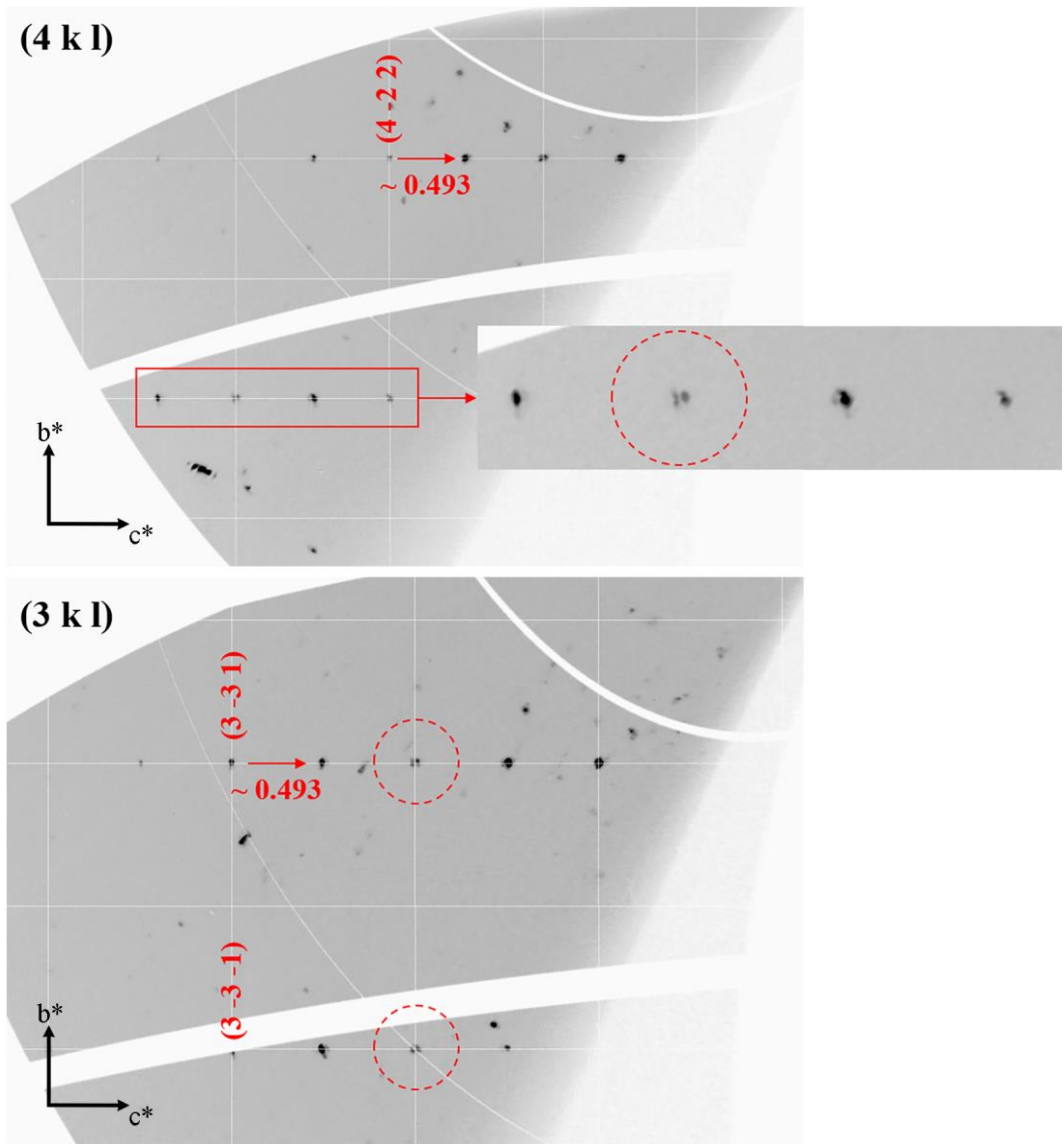

Fig. S3. Reconstructed reciprocal lattice planes of Br-II at 81(3) GPa. Red dashed circles highlight the splitting of the reflections. The details of the structural refinement are presented in Table. S3.

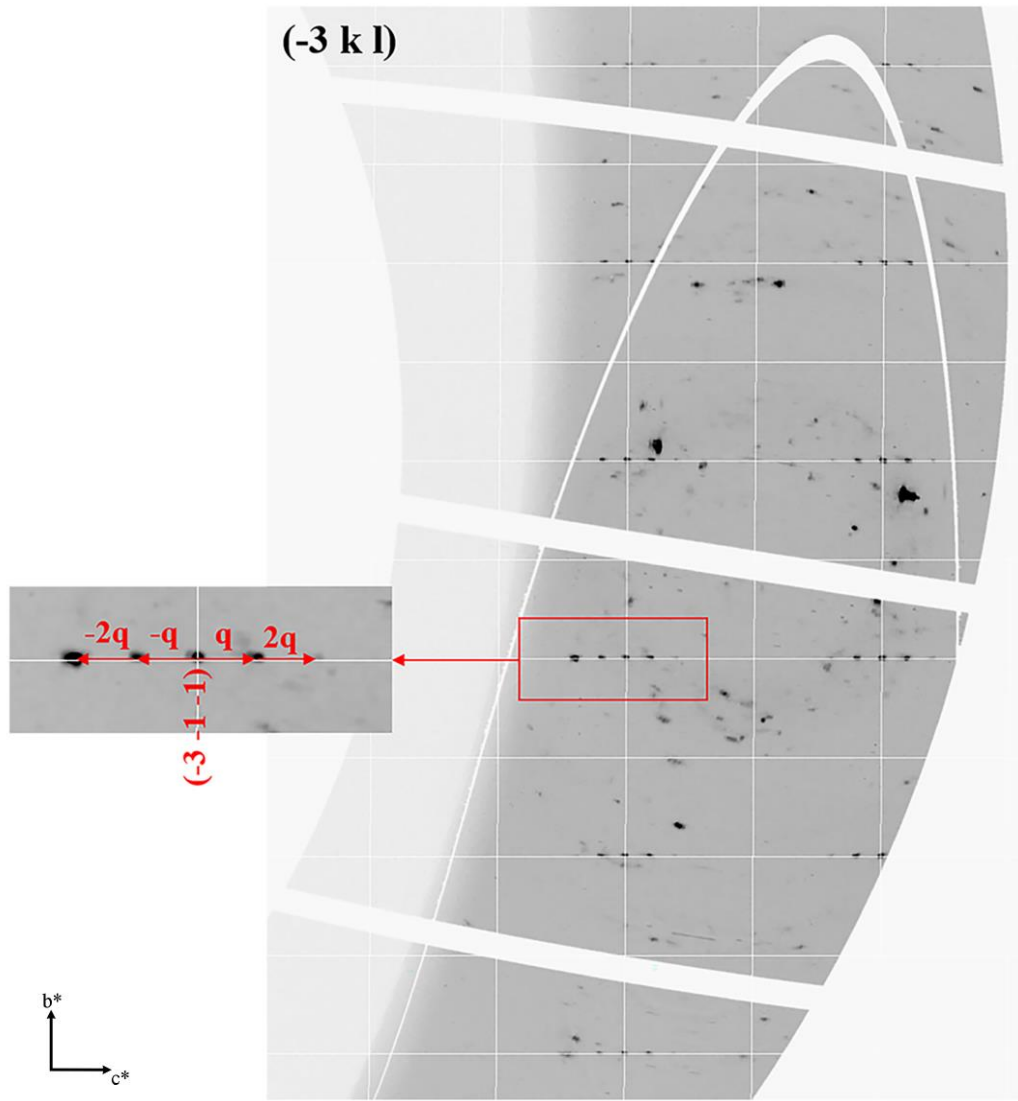

Fig. S4. Reconstructed precession image of the polytype Br-IIIγ (here γ = 0.1962(6), 96(3) GPa) in the *Fmmm* setting. Satellite reflections are indexed in terms of the modulation vector.

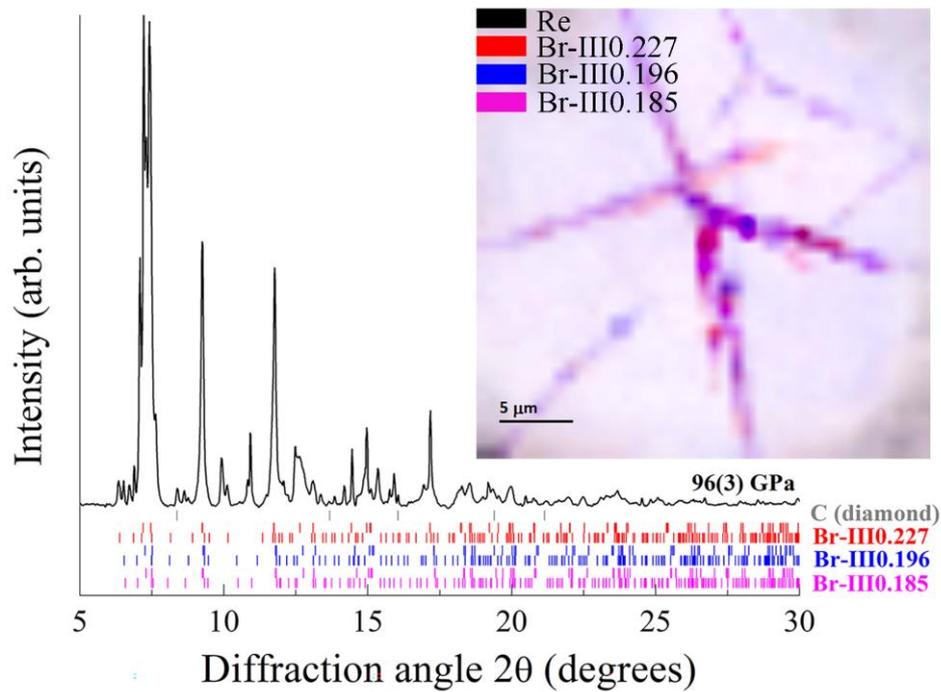

Fig. S5. Powder XRD pattern from the sample in DAC #1 revealing polytypism in solid bromine at 96(3) GPa after laser heating at ~ 2600 K. X-ray wavelength λ= 0.2846 Å, beam size of 0.75 × 0.75 μm². Three bromine polytypes (Br-IIIγ, γ ~0.227, ~0.196, and ~0.185) were identified in one spot using SCXRD data; the ticks of different colors correspond to calculated powder XRD patterns of these polytypes (red – Br-III0.227; blue – Br-III0.196; magenta – Br-III0.185); two rows of the same color indicate the positions of the main (upper ticks) and satellite (lower ticks) reflections of the same polytype. The inset is an XRD map of the whole sample chamber, which is identified on the reflections with indices (1 1 1 -2) and (1 1 1 -1) for Br-III0.227, (1 1 1 2) and (1 1 3 -2) for Br-III0.196, and (1 1 1 2) and (0 2 0 2) for Br-III0.185. It indicates a mixed distribution of the three polytypes within the sample.

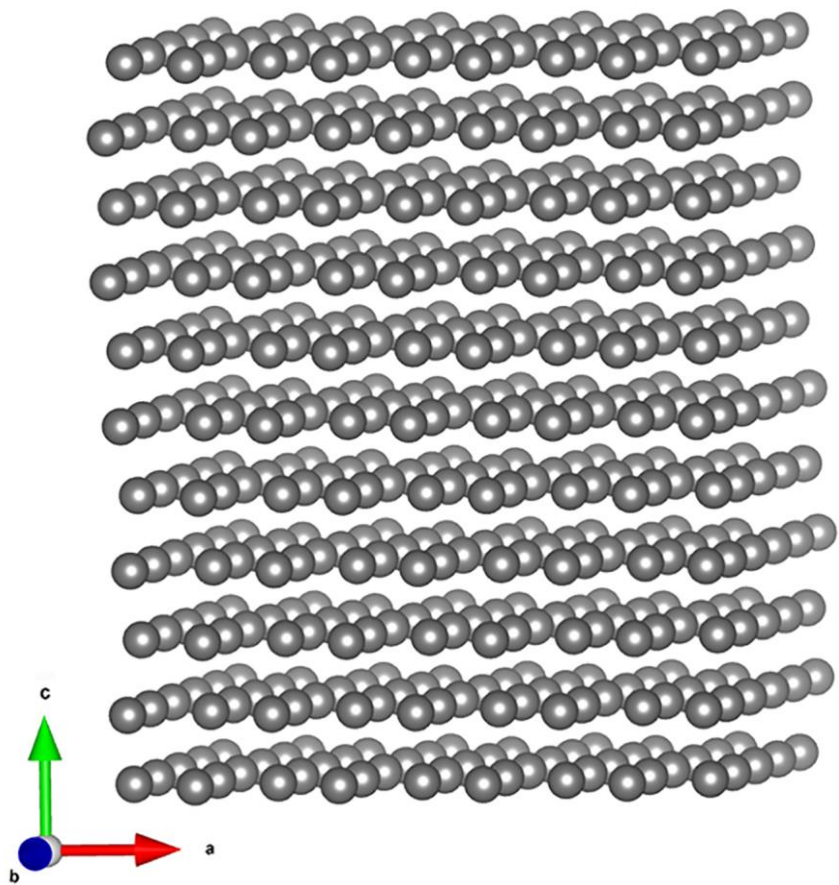

Fig. S6. A perspective view of the structure of Br-IIIγ, highlighting the layers of Br atoms in the *ab* plane stacked in the *c* direction. Br atoms are shown in grey balls.

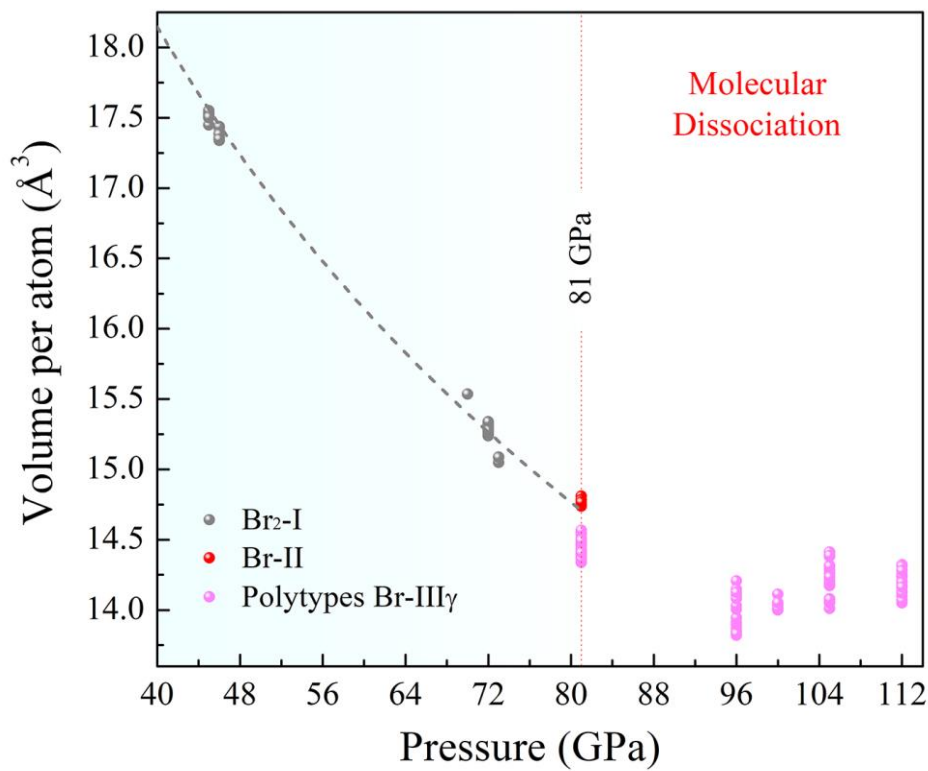

Fig. S7. Pressure dependence of the volume per atom of bromine at ambient temperature. The dashed gray line represents the fit of our experimental points for $Br_2$-I at 45, 46, 70, 72, and 73 GPa (see Supplementary Methods) using the 2$^{nd}$ order Birch-Murnaghan equation of state ($K_0$ = 22(3) GPa, $V_0$ = 32.10(15) Å$^3$/atom) (the volume per atom for $Br_2$-I at 1 atm. from Ref. [20] is also included to the fit). The vertical line indicates the apparent pressure boundary between molecular and non-molecular allotropes. The data are obtained from all domains which could be analyzed ($R_{int}$ < 10%). The uncertainties in the volume are smaller than the symbols' sizes. The uncertainty in pressure is 3 GPa.

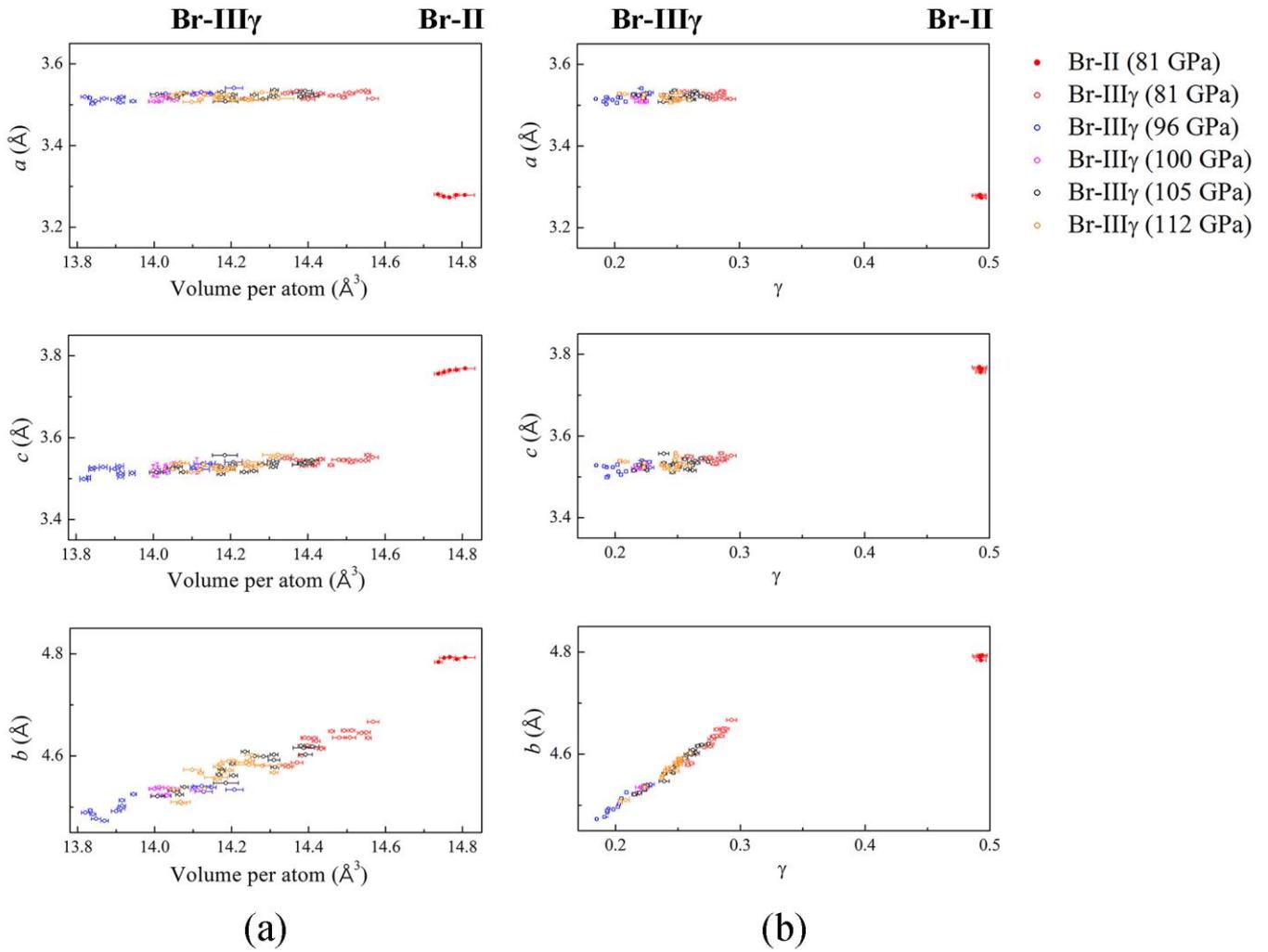

Fig. S8. Relations between lattice parameters and (a) volume per atom, and (b) the γ component for Br-II (filled symbols) and Br-IIIγ (open symbols). Data at 81(3), 96(3), 100(3), 105(3), and 112(3) GPa are shown in red, blue, magenta, black, and orange, respectively. For the *y*-axis, the uncertainty is smaller than the symbol size.

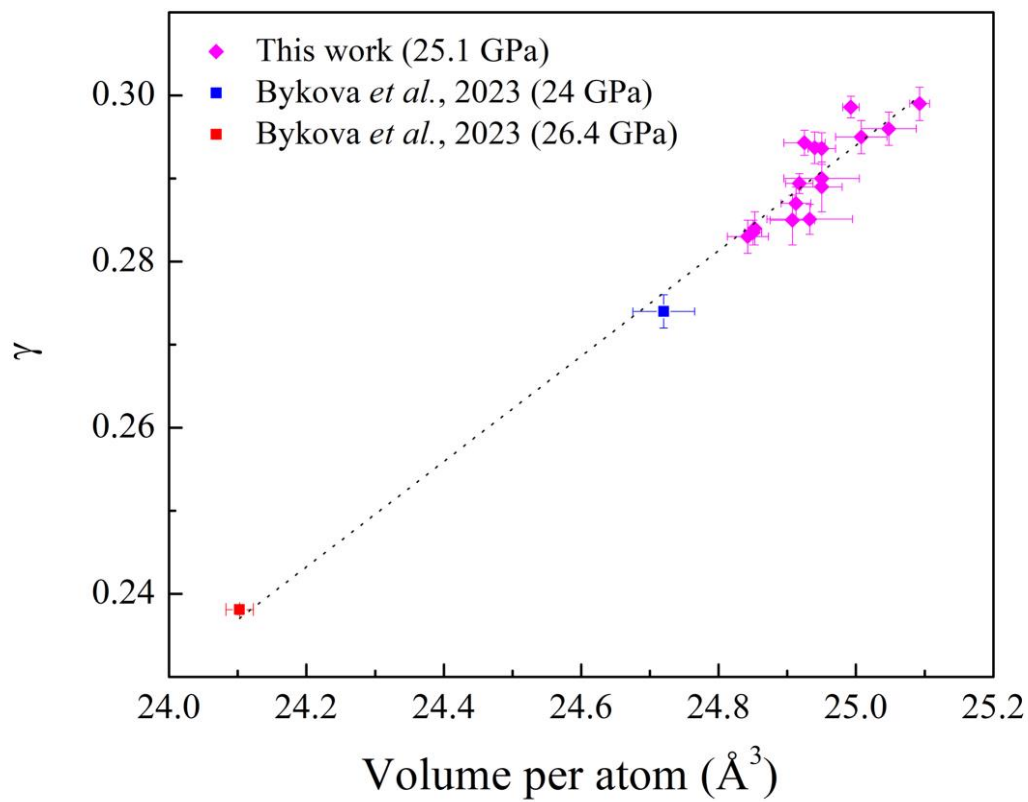

Fig. S9. The relationship between the values of γ and the volume per atom in the structure of iodine I-V; the dotted black line is a guide for the eyes. The results of this work (shown in magenta) are compared to previously reported data (blue and red) from Ref. [12].

# Supplementary References